\documentstyle[aps,prd,epsfig]{revtex}
\newcommand{\VEV}[1]{\left\langle #1\right\rangle}
\renewcommand{\Re}{\mathop{\rm Re}}	    
\renewcommand{\Im}{\mathop{\rm Im}}	    
\begin{document}
\draft
%
%
\title{
Potts Flux Tube Model at Nonzero Chemical Potential
}
\author{Jac Condella and Carleton DeTar}
\address{
Department of Physics, University of Utah, 
Salt Lake City, UT 84112, USA
}
%
%
\date{\today}
\maketitle
\begin{abstract}

We model the deconfinement phase transition in quantum chromodynamics
at nonzero baryon number density and large quark mass by extending the
flux tube model (three-state, three-dimensional Potts model) to
nonzero chemical potential.  In a direct numerical simulation we
confirm mean-field-theory predictions that the deconfinement
transition does not occur in a baryon-rich environment.

\end{abstract}
%
%
%
\section{Introduction}

Prospects of creating a quark-gluon plasma in the laboratory and
interest in the role of such a plasma in the early Universe and in
dense stars have stimulated efforts to understand the plasma/ordinary
matter phase transition starting from first principles in quantum
chromodynamics (QCD).  It is widely suspected that the phase
transition occurs at high density as well as at high
temperature. While much progress has been made in characterizing the
phase transition at high temperature at zero average baryon density in
full QCD, the same degree of success has not been achievable at
nonzero density, chiefly because the standard SU(3) lattice action
becomes complex, invalidating standard Monte Carlo methods.
Simulations must then be carried out with related ensembles of real,
positive weight.  Studies on small volumes offer intriguing hints
about the phase structure at nonzero density.  But with the wrong
ensemble, the thermodynamic limit cannot be taken, so the phase
structure cannot be ascertained
\cite{ref:review_zero,ref:review_nonzero}.  Thus we turn to simple
statistical models for insight.

The three-dimensional three-state Potts model is one of the standard
paradigms for lattice QCD in the strong-coupling, high-temperature,
large-quark-mass limit
\cite{ref:svetitsky_yaffe,ref:patel,ref:degrand_detar}.  It has been
used to provide qualitative information about the deconfinement phase
transition.  Past Potts model studies have been limited to
simulations at zero quark chemical potential and to mean-field
studies at nonzero chemical potential.  In this work we show that the
Potts model can be extended easily to nonzero baryon chemical
potential, permitting direct simulation using standard techniques.

The QCD phase transition changes character as the quark masses and
flavors are varied.  At zero quark mass with two or more flavors the
transition restores the spontaneously broken chiral symmetry
\cite{ref:pisarski_wilczek}.  At infinite quark mass it leads to
``deconfinement'' in the pure Yang-Mills theory.  Simulations of full
QCD with two flavors at zero chemical potential show that these two
regimes are separated at intermediate quark mass by a region where no
phase transition occurs---only a strong crossover.  In the
corresponding Potts model we find this obliteration of the
deconfinement transition, not only as the quark mass is lowered, but
also as the chemical potential is increased.  The former is found
from direct simulation, but the latter is known until now only in
mean field theory \cite{ref:degrand_detar}.  Here we confirm the mean
field prediction with a direct simulation.

Because the Potts model analogy works only in the large quark mass
limit, we learn only the fate of the deconfinement transition and not
the chiral transition.  Nonetheless, to the extent that a strong
crossover in QCD, observed in the intermediate quark-mass regime, is a
vestige of deconfinement, one may speculate that a weakening of the
QCD crossover at nonzero chemical potential is then indicated.

In the following section we describe the model.  In Sec.~3 we present
results of a simulation.  A summary and discussion is offered in the
final section.

\section{The Model}

We use a variant of the flux tube model of Patel \cite{ref:patel} also
discussed in \cite{ref:bardens}.  The model describes a classical
statistical system with no dynamics, consisting of a three-dimensional
cubic lattice with quarks and antiquarks occupying the sites and color
flux tubes occupying the nearest-neighbor links.  A configuration is
characterized by the quark number distribution $n_r \in [-3,-2,...,3]
$ and the color flux $\ell_{r,i} \in [-1,0,1]$ for each lattice site
$r$ and associated links in the $i = 1,2,3$ directions.  For
convenience links entering a site from a negative direction are
denoted alternatively by $\ell_{r,-i} = \ell_{r-\hat i,i}$.  Gauss's
law is enforced in modulo 3 arithmetic:
\begin{equation}
   \sum_{i = 1}^3 (\ell_{r,i} - \ell_{r,-i}) = n_r \ \ \ 
    \hbox{mod 3}.
\end{equation}
The hamiltonian assigns a mass $m$ to each quark and flux link energy
$\sigma$ to each link:
\begin{equation}
   H = \sum_{r,i}\sigma|\ell_{r,i}| + \sum_r m|n_r|
\end{equation}
The grand canonical partition function at inverse temperature $\beta$
and quark chemical potential $\mu$ is then
\begin{equation}
   Z_f(\beta,\mu) = \sum_{\{\ell_{r,i},n_r\}^\prime}
       \exp[-\beta (H - \mu N)]
 \label{eq:zf}
\end{equation}
where $N = \sum_r n_r$ and the prime indicates that the sum is over
all configurations satisfying Gauss's law.

The model is strictly static.  The lowest vacuum excitations consist
of mesons built from a quark-antiquark pair separated by one flux link
and baryons consisting of three quarks on a site.  Further excitations
lead to extended baryons and mesons and more complex hadrons, always
of zero triality.  (We limited quark occupation to a maximum of three
per site, but do not expect qualitative changes in our results if we
increase this limit.)

The flux tube model is equivalent to the three-state
three-dimensional Potts model with complex magnetic field.  That is,
$Z_f \propto Z_p$, where the Potts model partition function is
\begin{equation}
  Z_p = \sum_{z_r}\exp\left(\beta^\prime J\sum_{r,i}\Re(z_r z^*_{r+i})
    +  \beta^\prime h        \sum_r \Re z_r 
    + i\beta^\prime h^\prime \sum_r \Im z_r\right)
\end{equation}
This is the form of the Potts model found from the high-temperature,
high-quark-mass, strong-coupling limit of lattice QCD
\cite{ref:svetitsky_yaffe,ref:degrand_detar}.
The equivalence is established by a change of basis.  The derivation starts
by replacing the Gauss's law constraint at each site by a Kronecker
delta in mod 3:
\begin{equation}
  \frac{1}{3}\sum_{z \in Z(3)} z^\ell = \delta_{\ell,0}
\end{equation}
When we introduce one such sum for each site in the lattice, the sums
over link and site occupation numbers decouple and can be summed
explicitly.  The $Z(3)$ constraint variables $z_r$ become the
``clock'' spins of the Potts model.  After rearranging the sums we
have
\begin{eqnarray}
 \nonumber  Z_f(\beta,\mu) &=& \sum_{z_r}
     \prod_{r,i}\left(\sum_{\ell_{r,i}}
       \exp(-\beta\sigma |\ell_{\bf r,i}|) 
          (z_r z^*_{r+\hat i})^{\ell_{\bf r,i}}\right) \\
 & & \prod_r\left(\sum_{n_r}
       \exp[-\beta (m |n_{\bf r}| - \mu n_{\bf r})]
          z_r^{-n_r}\right)
\end{eqnarray}
The sums under the product symbols are explicitly
\begin{eqnarray}
   && 1 + 2 Re (z_{\bf r} z_{\bf r+\hat i}^*)\exp(-\beta\sigma) 
   \label{eq:factor1}\\ 
   && 1 + u^3 + v^3 + z(v^2 + u) + z^*(u^2 + v)
   \label{eq:factor2}
\end{eqnarray}
where 
\begin{equation}
  u = \exp[-\beta(m + \mu)] \ \ \mbox{and} \ \ \ v = \exp[-\beta(m - \mu)].
\end{equation}
With the aid of an identity over $Z(3)$ these factors can be rewritten
in exponential form.  The identity we need is
\begin{equation}
  1 + az + bz^* = \exp(c + d  \Re z + i e  \Im z)
\end{equation}
where
\begin{eqnarray}
  3c &=&  \ln(1 + a + b) + \ln(1-a-b+a^2+b^2-ab) \\
  3d &=& 2\ln(1 + a + b) - \ln(1-a-b+a^2+b^2-ab) \\
  \frac{\sqrt{3}}{2} e &=& \arctan\left(\frac{\sqrt{3}( a - b) }
           {2 - a - b }\right)
\end{eqnarray}
Applying this identity to the first factor (\ref{eq:factor1}) gives
the relation between the Potts spin coupling and the string energy (in
units of the respective temperatures)
\begin{equation}
  J\beta^\prime = \frac{2}{3}\ln\left(\frac{1+2\exp(-\beta\sigma)}
  {1-\exp(-\beta\sigma)}\right)  
\end{equation}
and to the second factor (\ref{eq:factor2}), gives the relation between
the Potts magnetic fields and the quark mass and chemical potential
\begin{eqnarray}
  h\beta^\prime        &=& \frac{2}{3}\ln(1 + a + b) 
   - \ln(1-a-b+a^2+b^2-ab) \\
  h^\prime\beta^\prime &=& \frac{2}{\sqrt{3}}
   \arctan\left(\frac{\sqrt{3}( a - b) }
           {2 - a - b }\right)
\end{eqnarray}
where
\begin{eqnarray}
  a &=& \frac{v^2 + u}{1 + u^3 + v^3} \\
  b &=& \frac{u^2 + v}{1 + u^3 + v^3}
\end{eqnarray}

Notice that $h^\prime$ is odd in $\mu$ and $h$ is even.  At zero
chemical potential the imaginary coupling vanishes ($h^\prime = 0$)
and at infinite quark mass the real field $h$ also vanishes.  In this
limit the Potts model exhibits the well known first order transition
at a value $\beta_t^\prime J = 0.36703(14)$, known from numerical
simulation
\cite{ref:potts}.  Here we have the Potts/flux-tube parameter
equivalence
\begin{equation}
(\beta_t^\prime J = 0.3670,
 \beta_t^\prime h = \beta_t^\prime h^\prime = 0) \equiv 
  (\beta\sigma = 1.6265, \beta m = \infty, \beta \mu = 0)
\end{equation}
Numerical simulation has indicated that the first order transition
persists to a small real magnetic field \cite{ref:degrand_detar}.  The
critical endpoint is crudely known to be in the range $[0.002,0.01]$
for which
\begin{equation}
(\beta^\prime J = 0.365, \beta^\prime h = [0.002,0.01], \beta^\prime h^\prime = 0)
 \equiv 
(\beta \sigma = 1.632, \beta m = [3.2,4.2], \beta \mu = 0) .
\end{equation}
Any mass $m$ higher than this critical value should admit a first
order phase transition.

We chose to simulate the flux tube model at quark mass $m/\sigma = 5$.
This value was selected to assure a first order phase transition at
zero chemical potential.  Figure~\ref{fig:potts_map_z_mu} shows the
mapping from flux tube $\beta\sigma$ to Potts parameters at zero
chemical potential.  We then repeated the simulation at nonzero
chemical potential $\mu/\sigma = 1.75$.  The corresponding map is
shown in Fig.~\ref{fig:potts_map_nz_mu}.  The Potts magnetic fields
remain small.

\section{Method and Results}

We carry out a simulation in the occupation number basis
(\ref{eq:zf}), {\it i.e.} the flux tube formulation, in which the Boltzmann
weights are real and positive and the simulation can be done easily
using conventional Metropolis methods.  We have chosen an elementary
set of Metropolis moves that preserve the Gauss's law constraint, and
are capable of reaching any valid configuration.  The local moves
consist of systematically ``adding'' or ``subtracting'' one of four
elementary color-singlet hadrons at all locations and orientations in
the configuration.  A single sweep of the lattice consists of
considering each of these moves for all orientations of the hadrons at
each lattice site in typewriter order. 

The Metropolis-move hadrons are these: (1) a quark and antiquark
separated by one flux link, (2) a diquark and antidiquark separated by
one flux link, (3) a quark and diquark or antiquark and antidiquark,
also separated by one flux link, (4) and a plaquette of flux links.
The process of adding (or subtracting) a hadron consists of
increasing or decreasing the quark occupation number and flux link
value of the configuration in mod 3 arithmetic according to the
position and orientation of the hadron selected, respecting our
exclusion principle that limits the quark number to the range
$[-3,3]$.  By always adding or subtracting a color singlet state,
Gauss's law is always obeyed.  While all configurations satisfying
Gauss's law can be reached by a combination of these moves, this
over-complete set was also chosen in an effort to cover the phase
space efficiently.  Still, we have only a local algorithm, presumably
as effective as a local algorithm in the spin basis.  For the moment
we have not considered cluster algorithms analogous to those that have
been so successful in spin systems \cite{ref:cluster}.

As we have mentioned we simulate at fixed quark mass $m/\sigma = 5$
and choose two values of the chemical potential, namely $\mu/\sigma =
0$ and $1.75$.  We then vary $\beta \sigma$ to locate the phase
transition or crossover.  We expect to reproduce the Potts first order
phase transition at zero chemical potential and small real field, but
at nonzero chemical potential we explore new territory.  We simulate
at a series of volumes $L^3$ for $L = 10, 20, 30, 40$ in each case.
In the crossover region we extend the simulations for, typically,
35,000 to 60,000 Metropolis sweeps.

As expected at zero chemical potential, we find a sharp rise in energy
density at $\beta\sigma \approx 1.63$ as shown in
Fig.~\ref{fig:energy_z_mu}.  To characterize the phase transition we
study the size dependence of the peak in specific heat
\cite{ref:fss_first_order}.  The specific heat is defined in the usual
way in terms of the total energy $E$ of the configuration:
\begin{equation}
    C_V/\beta^2 = (\langle E^2 \rangle - \langle E \rangle^2)/L^3.
\end{equation}

Figure~\ref{fig:spheat_z_mu} shows a peak in the specific heat at zero
chemical potential that sharpens and grows with increasing volume.
Also shown in these figures are results of a fit to a
phenomenological form for a first order phase transition based on a
simplification of the finite size analysis of Borgs {\em et al.}
\cite{ref:Borgs}:
\begin{equation}
   C_V/\beta^2 = \frac{C_{V,\rm max}(L)}
   {\cosh^2[\gamma(L)(\beta - \beta_c(L))]}.
\end{equation}
The simplification assumes that the probability distribution for the
energy density near the phase transition receives a delta function
contribution from the two phases with the usual Boltzmann weights:
\begin{equation}
  P(E) \propto \delta(E - E_d)\exp(-\beta E_d + S_d) +
    \delta(E - E_o)\exp(-\beta E_o + S_o)
\end{equation}
in terms of the energy and entropy in the ordered and disordered
phases.  This model gives
\begin{eqnarray}
  \gamma &=& L^3 Q/2 \nonumber \\
  C_{V,\rm max} &=& L^3 Q^2/4 \label{eq:scale}
\end{eqnarray}
for $Q = (E_d - E_o)/L^3$, the latent heat of the transition.  In fact
the actual probability distribution resembles a sum of broad peaks,
but since we are concerned only with extracting $C_{V, \rm max}$, it
suits our purpose.  Although there are really only two independent
parameters, $\beta_c$ and $Q$, we have kept the height, width, and
center unconstrained and look for evidence for scaling with $L^3$.

In Fig.~\ref{fig:spheatmax_z_mu} we plot the specific heat maximum
{\it vs} $L^3$ and show that a linear relationship is plausible for $L
\ge 20$.  Errors are obtained using a bootstrap method, averaging
simulation results in blocks, and extrapolating to infinite block
size.  Departures from linearity for smaller $L$ are well known in
this model and arise from finite size effects.  We take the
approximate scaling of the peak in specific heat as good evidence for
a first order phase transition.

The quark number susceptibility (three quarks = one baryon), defined
as
\begin{equation}
   \chi_q = \frac{dB}{d\mu} = 9(\langle B^2 \rangle - \langle B \rangle^2)/L^3
\end{equation}
with $B = \VEV{\sum_r n_r/3}$, the total baryon number, rises as the
temperature is increased past the phase transition, as shown in
Fig.~\ref{fig:barsusc_z_mu}.

Having tested the method, we turn to nonzero chemical potential.
Shown in Fig.~\ref{fig:energy_nz_mu} is a plot of the energy density
{\it vs} $\beta\sigma$.  We see evidence for a weak crossover.  Notice
that here and in the remaining figures, we have enlarged the
$\beta\sigma$ scale compared with the corresponding zero chemical
potential figures.  Thus the crossover is far less abrupt than at zero
chemical potential. The corresponding specific heat is shown in
Fig.~\ref{fig:spheat_nz_mu}.  There is a diffuse peak in specific
heat, but we see no evidence that the peak height is changing with
increasing volume.  The peak height is considerably smaller than at
zero chemical potential, consistent with a broadening of the
crossover.  Thus we find no evidence of a phase transition at this
chemical potential and quark mass.

We turn to the quark number density and susceptibility.  Figure
\ref{fig:bardens_nz_mu} shows the quark number density as a function
of $\beta\sigma$.  We note that in the high temperature range, the
quark number density at our chosen mass is quite low -- namely, only a
few per thousand sites.  The density rises as temperature is increased
past the crossover.  As with the energy density, with the exception of
the $10^3$ volume, we see no evidence for a sharpening of the
crossover with increasing volume.  The quark number susceptibility
also shows no apparent trend with increasing volume, as shown in
Fig.~\ref{fig:barsusc_nz_mu}.


\section{Summary and Discussion}

We have extended the equivalence between the QCD-like flux tube model
and Potts model to nonzero chemical potential, making possible a
direct numerical simulation using standard Monte Carlo methods, where
only mean field methods have previously succeeded.  We have developed
numerical evidence that confirms the predictions of mean field theory,
namely that the deconfining phase transition disappears if the baryon
density is slightly nonzero.  While our results are not applicable in
the most interesting limit of small quark mass, one may speculate,
nonetheless, that they indicate a weakening of the crossover at high
baryon number density.

A recent study by Blum, Hetrick and Toussaint analyzes the static
limit of QCD and arrives at a similar conclusion concerning the fate
of the deconfining phase transition at high density \cite{ref:BHT}.
Their resulting static action is quite similar to the Potts model
action with imaginary magnetic field, so their method is also limited
to small volumes in a direct Monte Carlo simulation.  To obtain the
comparable limit in the flux tube model, one takes $m \rightarrow
\infty$ with $m - \mu$ fixed.  With only two variables, namely
$\beta\sigma$ and $\beta(m - \mu)$, one obtains a two-dimensional
subspace of the Potts parameter space $(J\beta^\prime, h\beta^\prime,
h^\prime\beta^\prime)$.  For large $m-\mu$ one has $h = h^\prime
\approx 0$, where the first order deconfining transition occurs,
whereas at small $m-\mu$ one explores the region of large
$h\beta^\prime$ and large $h^\prime\beta^\prime$.  In this way the
results can be compared.

In the corresponding canonical-ensemble study in the static limit of
QCD, a weakening of the crossover signal is also found at nonzero
baryon number \cite{ref:Engels:1999tz}.  However, the authors
interpret their results as suggesting the coexistence of two phases in
the transition region.

Could a similar change of basis also render full QCD simulations
tractable at nonzero chemical potential?  Normally the QCD path
integral is formulated on a basis in which the vector potential is
diagonal and the fermion integration is completed explicitly.  The
analogous change of basis diagonalizes the color electric flux and
treats fermions in the Fock space basis, summing over states that are
color singlets in quark and gluon content.  At finite $N_c$ it appears
that the complexities of enforcing color neutrality in such a basis
are prohibitive.

\section*{Acknowledgements} 

We thank Doug Toussaint for helpful comments.  Some computations were
carried out on the IBM SP at the Utah Center for High Performance
Computing.


%
\figure{
 \epsfig{bbllx=0,bblly=0,bburx=840,bbury=840,clip=,
         file=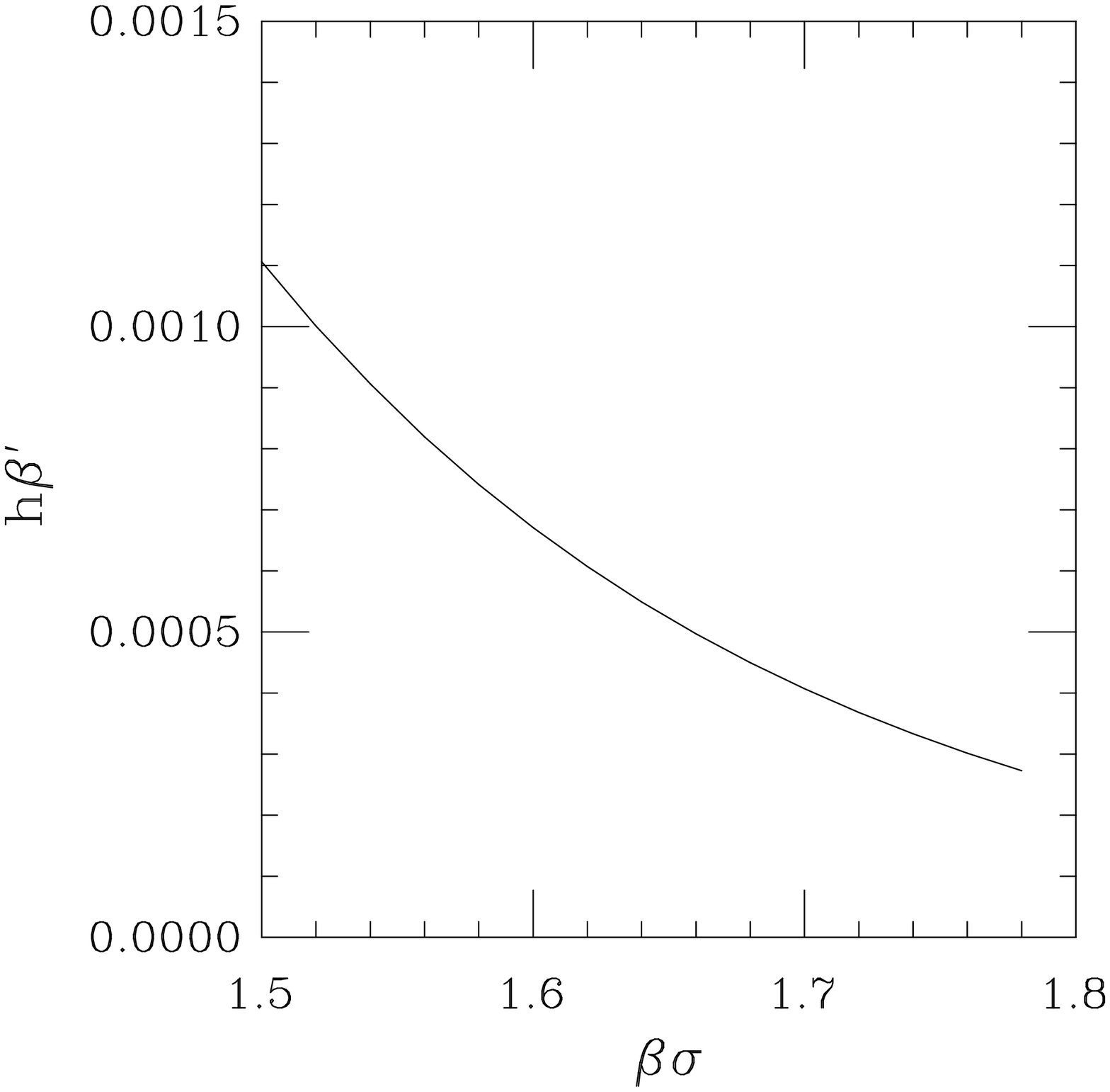,width=100mm}
\caption{Mapping of flux tube parameters $\beta\sigma$ to real Potts
magnetic field $\beta^\prime h$ at fixed quark mass $m/\sigma = 5$ and
zero chemical potential.  A first order phase transition is expected
at $\beta\sigma \approx 1.63$.
\label{fig:potts_map_z_mu}
}
}
\figure{
 \epsfig{bbllx=0,bblly=0,bburx=840,bbury=840,clip=,
         file=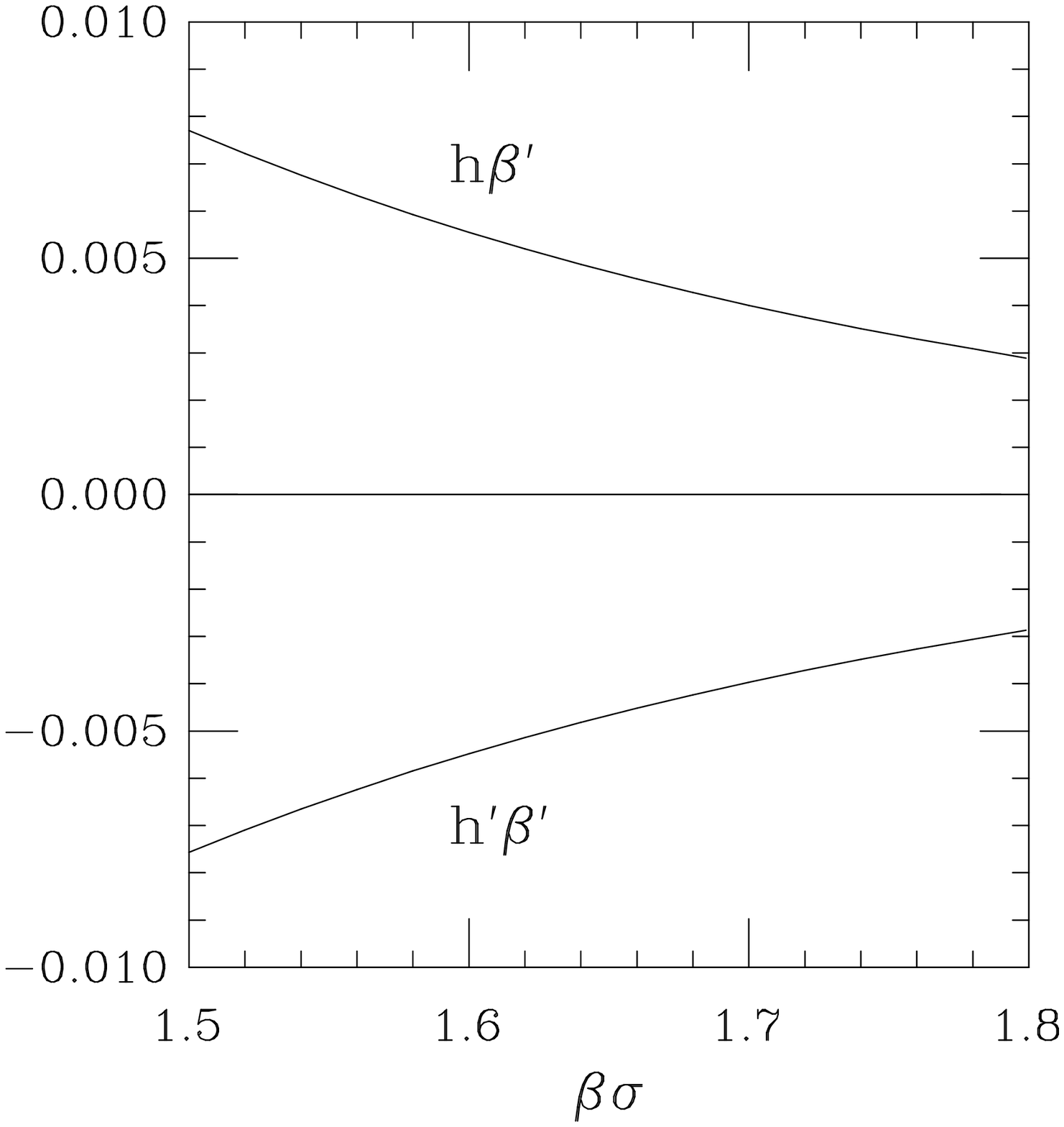,width=100mm}
\caption{Mapping of flux tube parameters $\beta\sigma$ to Potts real
and imaginary magnetic fields $\beta^\prime h$ and $\beta^\prime
h^\prime$ at fixed quark mass $m/\sigma = 5$ and nonzero chemical
potential $\mu/\sigma = 1.75$.
\label{fig:potts_map_nz_mu}
}
}
%
%
\figure{
 \epsfig{bbllx=200,bblly=130,bburx=830,bbury=940,clip=,
         file=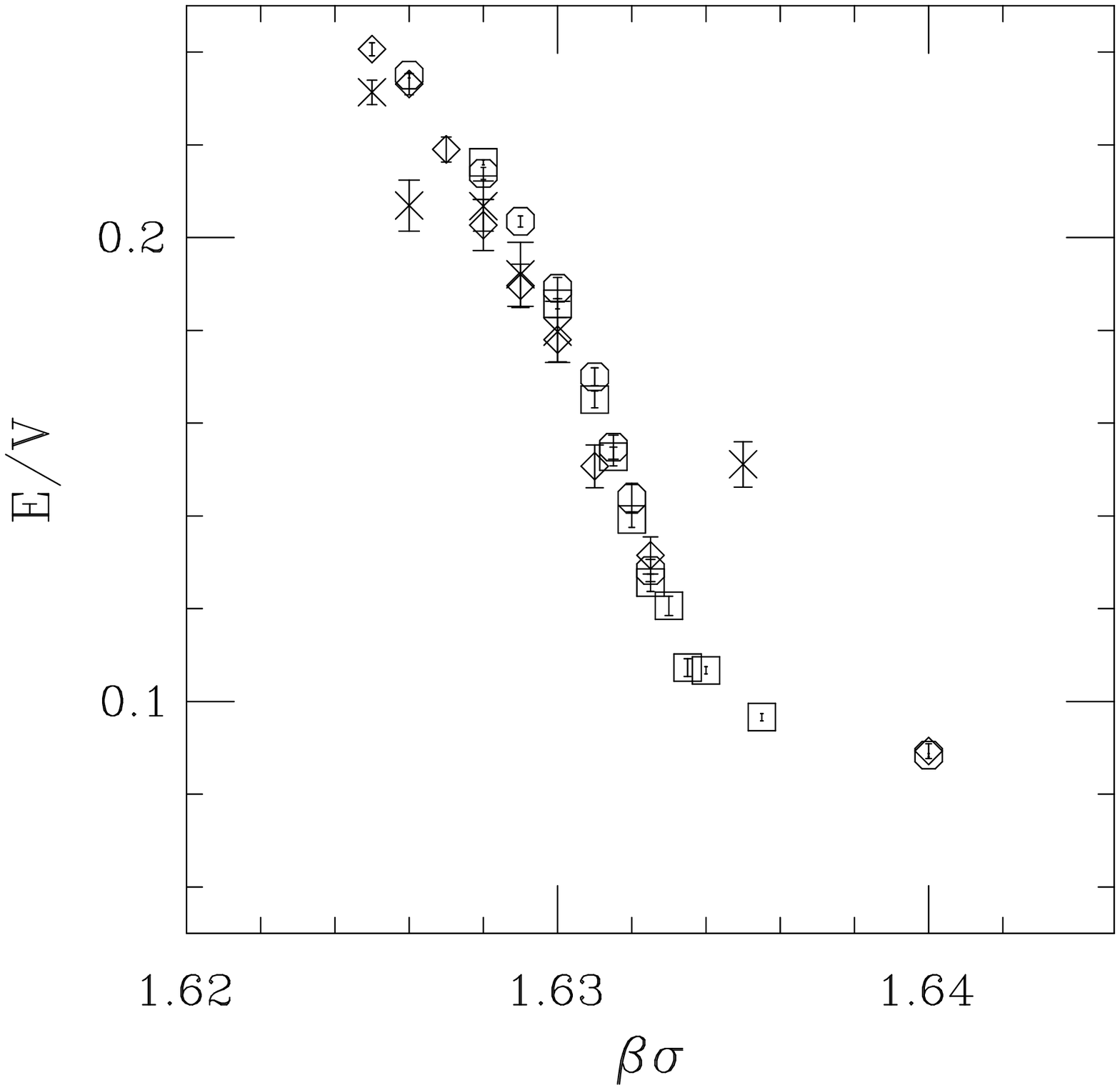,width=160mm}
\caption{Energy density {\it vs} $\beta\sigma$ in the critical region
at zero chemical potential for various lattice sizes: $L = 10$,
crosses; 20, diamonds; 30, octagons; 40, squares.
\label{fig:energy_z_mu}
}
}
\figure{
 \epsfig{bbllx=200,bblly=130,bburx=830,bbury=940,clip=,
         file=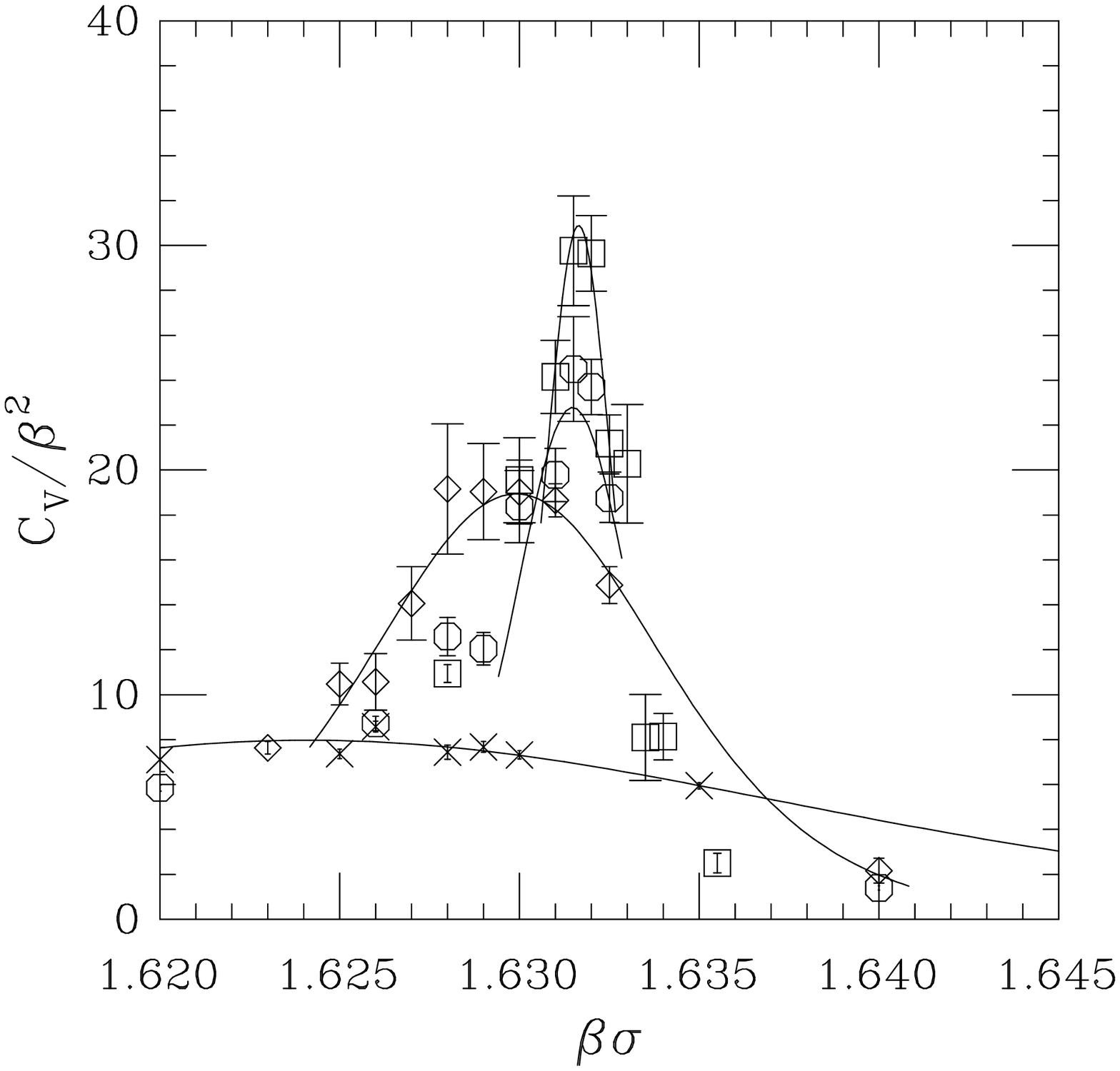,width=160mm}
\caption{Specific heat {\it vs} $\beta\sigma$ at zero chemical
potential for various lattice sizes.  Symbols as in
 Fig.~\protect\ref{fig:energy_z_mu}.
\label{fig:spheat_z_mu}
}
}
\figure{
 \epsfig{bbllx=200,bblly=130,bburx=830,bbury=940,clip=,
         file=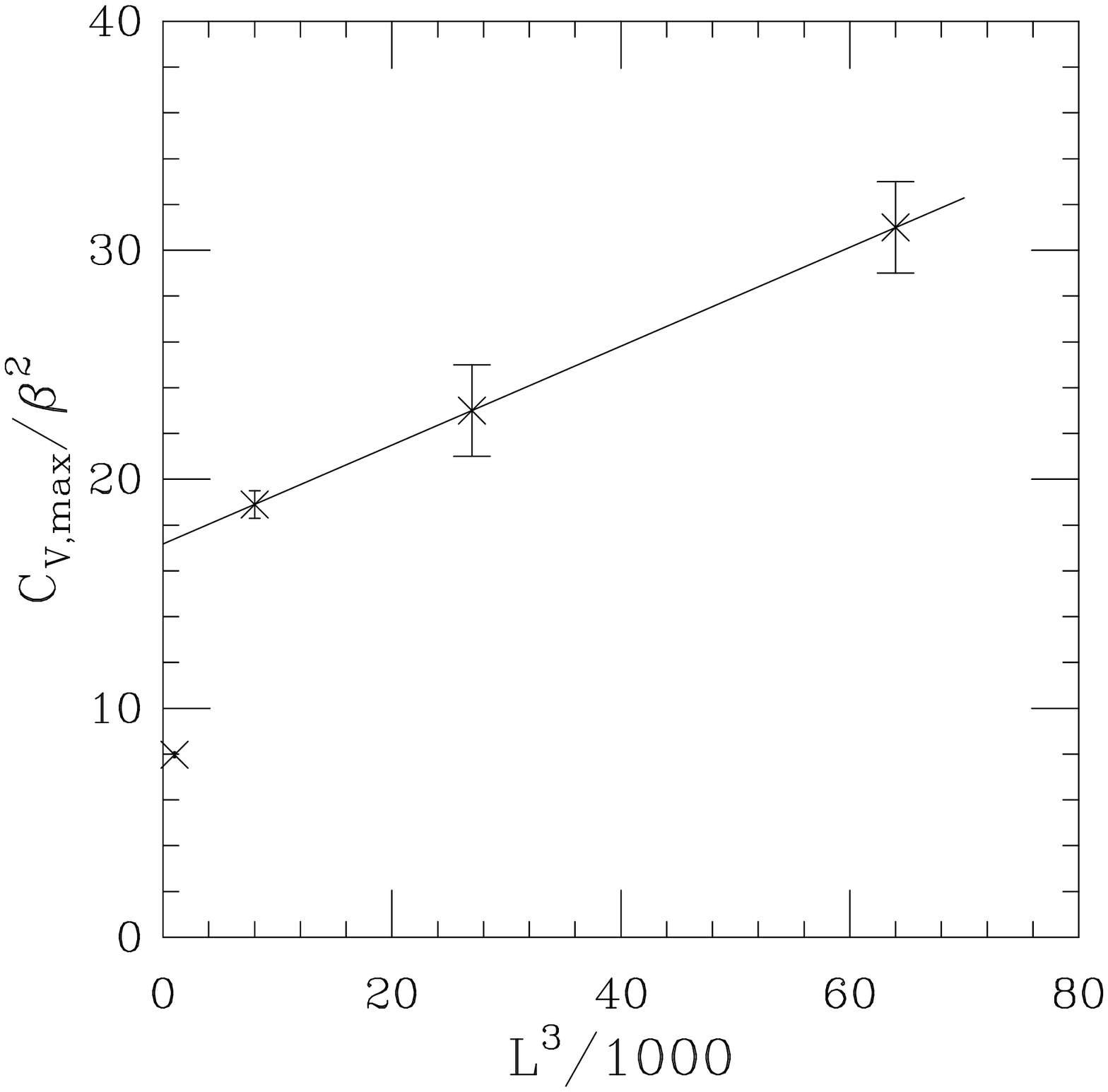,width=160mm}
\caption{Peak in specific heat {\it vs} lattice volume at zero
chemical potential
\label{fig:spheatmax_z_mu}
}
}
\figure{
 \epsfig{bbllx=200,bblly=130,bburx=830,bbury=940,clip=,
         file=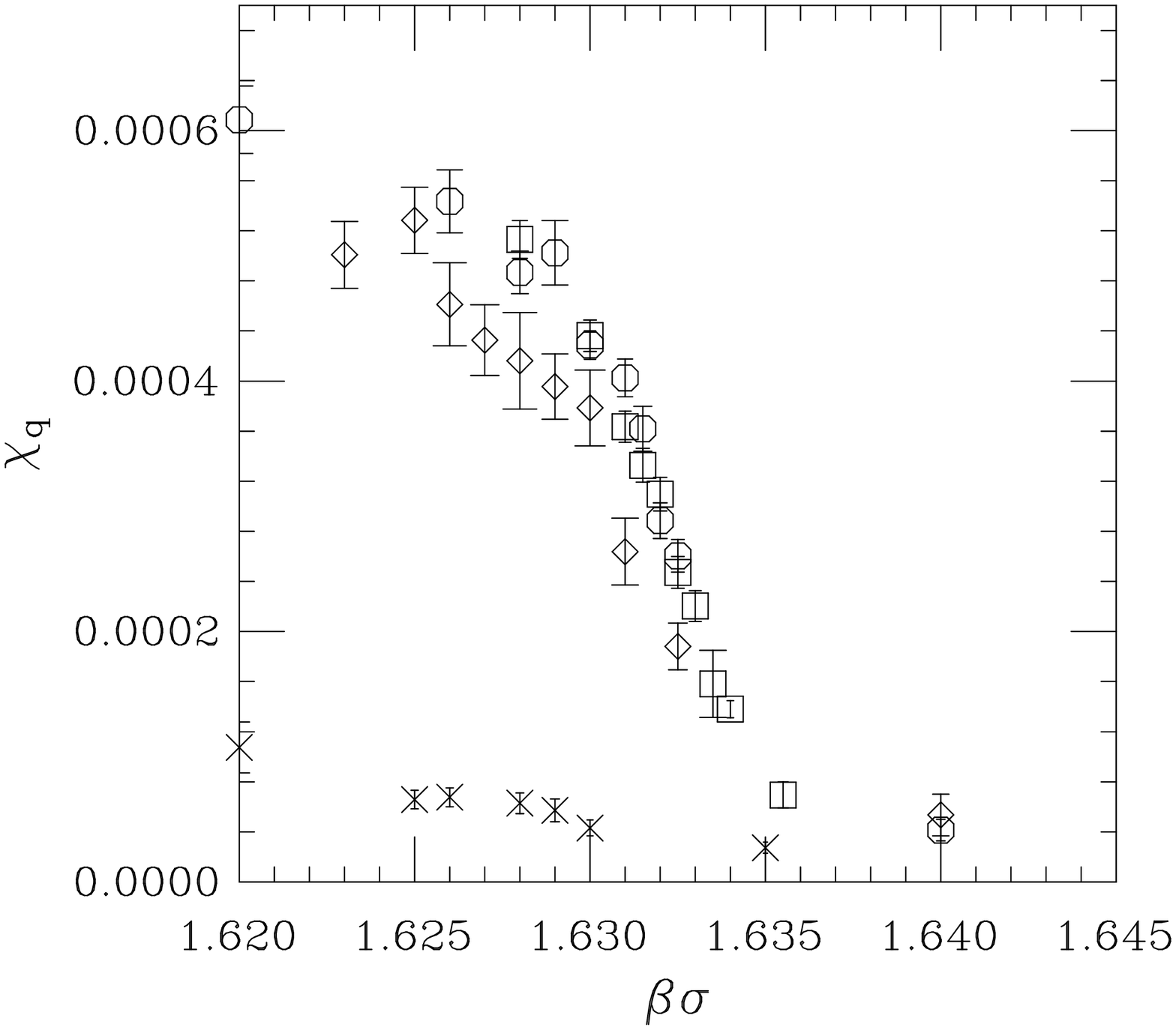,width=160mm}
\caption{Quark number susceptibility {\it vs} $\beta\sigma$ at zero
chemical potential for various lattice sizes.  Symbols as in
 Fig.~\protect\ref{fig:energy_z_mu}.
\label{fig:barsusc_z_mu}
}
}
\figure{
 \epsfig{bbllx=200,bblly=130,bburx=830,bbury=940,clip=,
         file=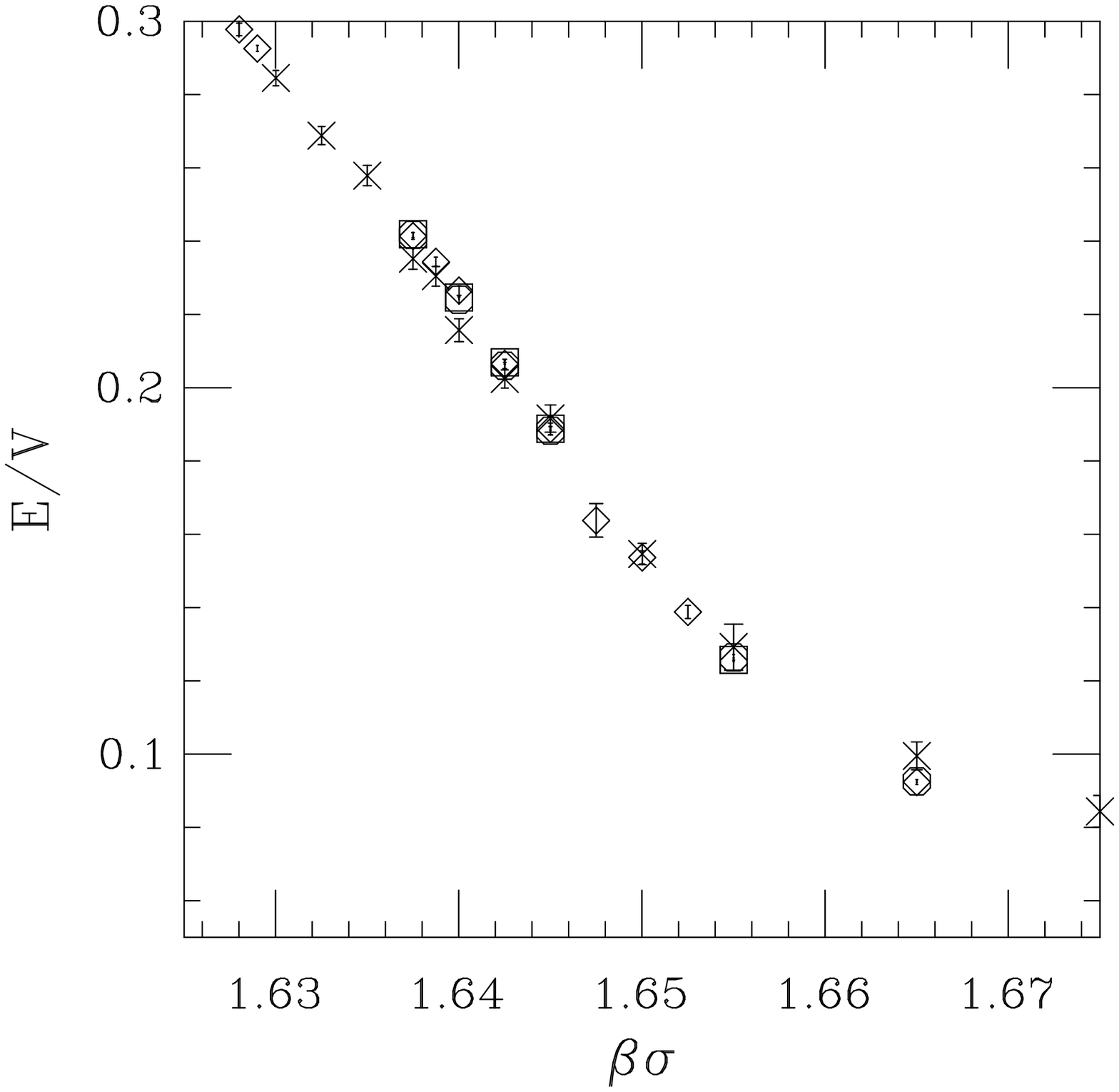,width=160mm}
\caption{Energy density {\it vs} $\beta\sigma$ at chemical potential
$\beta\mu = 1.75$ for various lattice sizes.  Symbols as in
 Fig.~\protect\ref{fig:energy_z_mu}.
\label{fig:energy_nz_mu}
}
}
\figure{
 \epsfig{bbllx=200,bblly=130,bburx=830,bbury=940,clip=,
         file=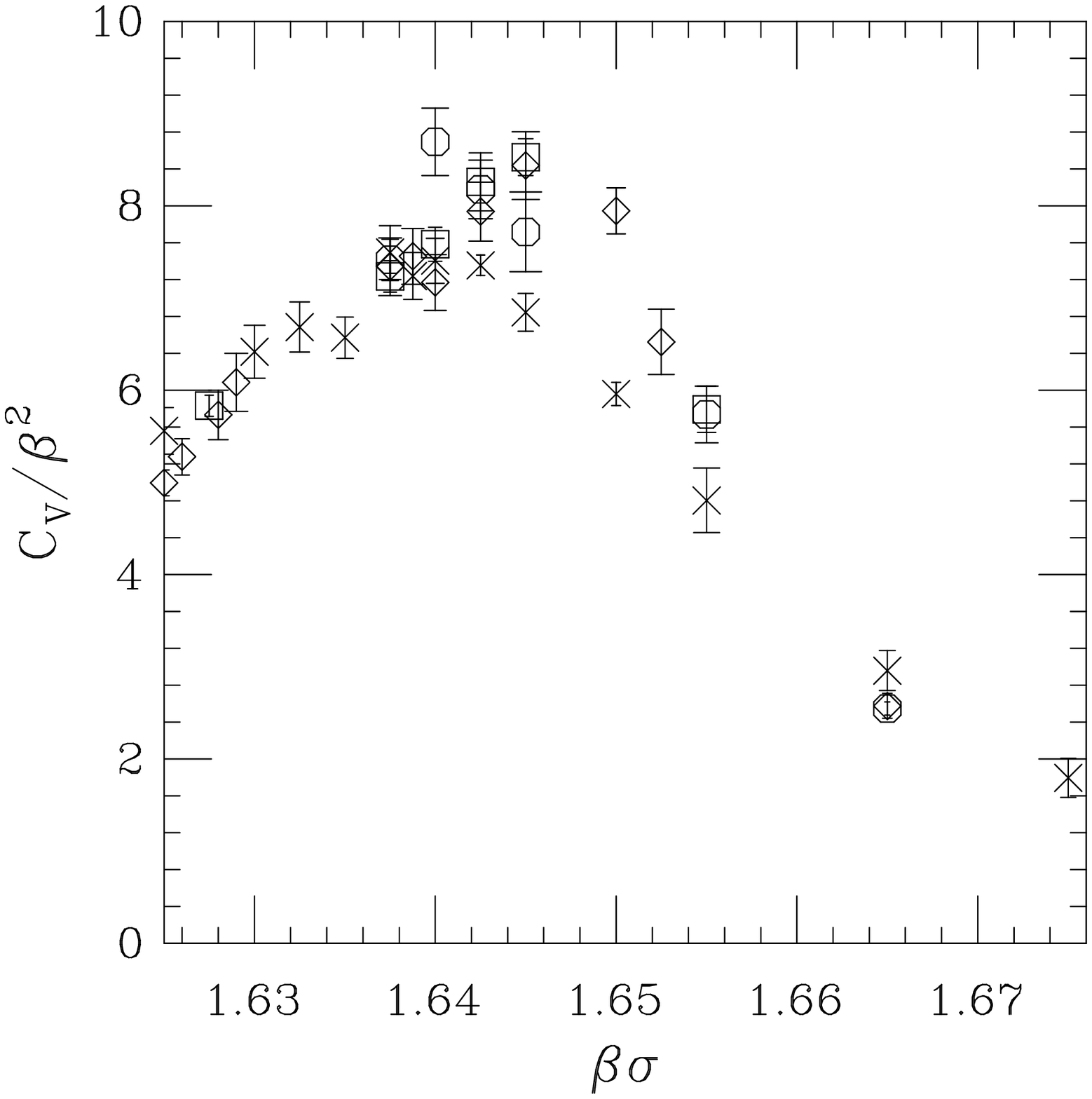,width=160mm}
\caption{Specific heat {\it vs} $\beta\sigma$ at chemical potential
$\beta\mu = 1.75$ for various lattice sizes.  Symbols as in
 Fig.~\protect\ref{fig:energy_z_mu}.
\label{fig:spheat_nz_mu}
}
}
\figure{
 \epsfig{bbllx=200,bblly=130,bburx=830,bbury=940,clip=,
         file=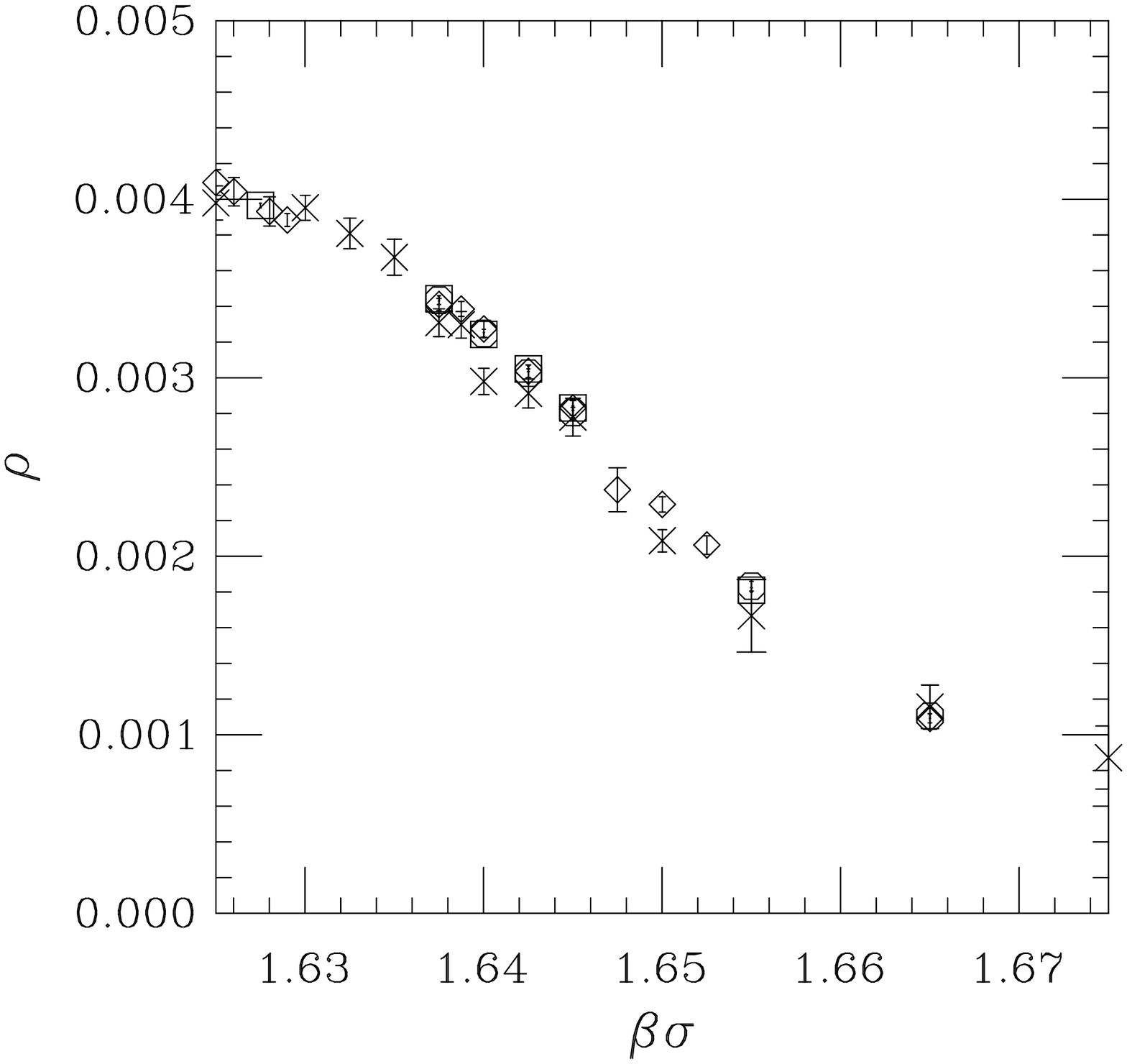,width=160mm}
\caption{Quark number density {\it vs} $\beta\sigma$ at chemical
potential $\beta\mu = 1.75$ for various lattice sizes.  Symbols as in
Fig.~\protect\ref{fig:energy_z_mu}.
\label{fig:bardens_nz_mu}
}
}
\figure{
 \epsfig{bbllx=200,bblly=130,bburx=830,bbury=940,clip=,
         file=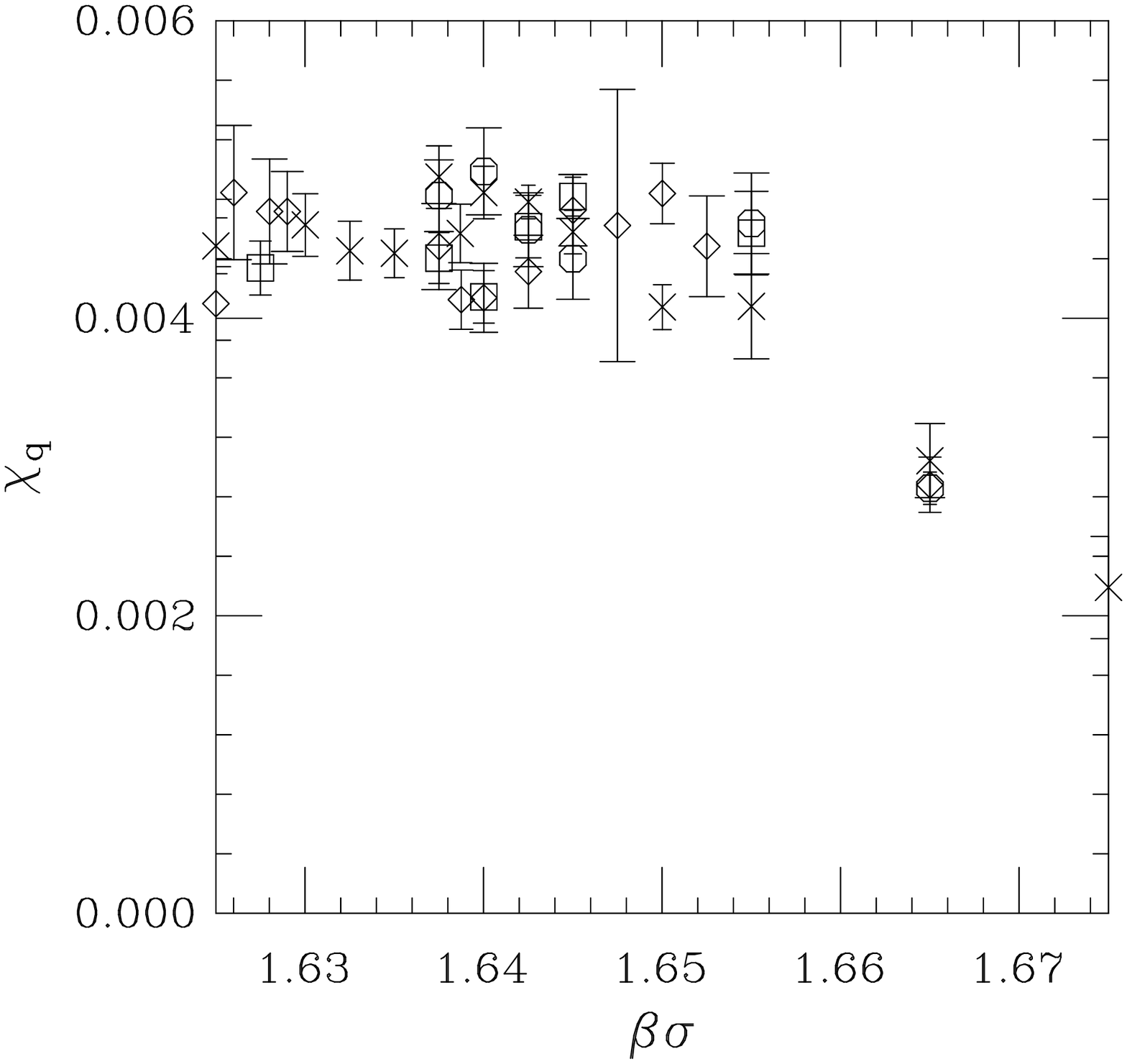,width=160mm}
\caption{Quark number susceptibility {\it vs} $\beta\sigma$ at chemical
potential $\beta\mu = 1.75$ for various lattice sizes.  Symbols as in
Fig.~\protect\ref{fig:energy_z_mu}.
\label{fig:barsusc_nz_mu}
}
}

\begin{references} 
%
\bibitem{ref:review_zero}
C.~DeTar, Nucl.\ Phys.\ B (PS) {\bf 42}, 73 (1995);
C.~DeTar in {\it Quark Gluon Plasma 2}, ed.\ R.~Hwa (World Scientific,
Singapore, 1995).
%
\bibitem{ref:review_nonzero}
For a brief review, see I.~Barbour, S.~Morrison, E.~Klepfish,
J.~Kogut, and M.~Lombardo, 
Nucl.\ Phys.\ B (Proc.\ Suppl.) {\bf 26}, 220 (1998).
%
\bibitem{ref:svetitsky_yaffe}
L.G.~Yaffe and B.~Svetitsky, 
Phys.\ Rev.\ D {\bf 26}, 963  (1982); 
A.M. Polyakov, 
Phys.\ Lett.\ B {\bf 72}, 477 (1978); 
L.~Susskind, 
Phys.\ Rev.\ D {\bf 20}, 2610 (1979).
%
\bibitem{ref:patel}
A.~Patel, 
Nucl.\ Phys.\ B {\bf 243}, 411 (1984); 
Phys.\ Lett.\ {\bf 139B}, 394 (1984).
%
\bibitem{ref:degrand_detar}
T.A.~DeGrand and C.~DeTar,
Nucl.\ Phys.\ B{\bf 225}, 590 (1983).
%
\bibitem{ref:pisarski_wilczek}
R.D.~Pisarski and F.~Wilczek,
Phys.\ Rev.\ D {\bf 46}, 4657 (1992);
F.~Wilczek,
J.\ Mod.\ Phys.\ {\bf A7}, 3911 (1992);
K.~Rajagopal and F.~Wilczek, 
Nucl.\ Phys.\ {\bf B399}, 395 (1993);
K.~Rajagopal in {\it Quark Gluon Plasma 2}, 
ed.\ R.~Hwa (World Scientific, Singapore, 1995).
%
\bibitem{ref:bardens}
C.~Bernard {\it et al.},
Phys.\ Rev. D {\bf 49}, 6051 (1994).
%
\bibitem{ref:potts}
W.~Janke and R.~Villanova, Nucl.\ Phys.\ B {\bf 489}, 679 (1997).
  R.V.\ Gavai, F.\ Karsch, B.\ Petersson, Nucl.\ Phys.\ B {\bf 322},
738 (1989); R.V.\ Gavai, S.\ Gupta, A.\ Irb\"ack, F.\ Karsch, B.\
Petersson, Nucl.\ Phys.\ B {\bf 329}, 263 (1990).
%
\bibitem{ref:cluster}
 R.H.~Swendson and J.-S.~Wang, Phys.\ Rev.\ Lett.\ {\bf 58}, 86
(1987).
%
\bibitem{ref:fss_first_order}
R.V.\ Gavai, F.\ Karsch, B.\ Petersson, Nucl.\ Phys.\ B {\bf 322},
738 (1989); R.V.\ Gavai, S.\ Gupta, A.\ Irb\"ack, F.\ Karsch, B.\
Petersson, Nucl.\ Phys.\ B {\bf 329}, 263 (1990).
%
\bibitem{ref:Borgs}
C.~Borgs, R.~Koteck\'{y}, and S.~Miracle-Sol\'{e}, 
J.~Stat.~Physics {\bf 62},  529  (1991).
%
\bibitem{ref:BHT}
T.~Blum, J.~Hetrick, and D.~Toussaint, Phys.\ Rev.\ Lett.\ {\bf 76},
1019 (1996).
%
\bibitem{ref:Engels:1999tz}
O.~Kaczmarek, J.~Engels, F.~Karsch and E.~Laermann,
``Lattice QCD at nonzero baryon number,''
hep-lat/9905022;
O.~Kaczmarek, F.~Karsch, E.~Laermann and M.~Lutgemeier,
``Heavy quark potentials in quenched QCD at high temperature,''
hep-lat/9908010;
J.~Engels, O.~Kaczmarek, F.~Karsch and E.~Laermann,
17th International Symposium of Lattice Field Theory ``Lattice 99'', 
Pisa, Italy, hep-lat/9908046.
%
\end{references}
\end{document}